\documentclass[useAMS,usenatbib,amsmath,amssymb,referee]{mn2e}
\usepackage{graphicx}
\usepackage{bm}
\usepackage{longtable}
\usepackage{dcolumn}
\usepackage{amsmath}
\usepackage{amssymb}
\usepackage{hyperref}

\usepackage{multirow}

\input{epsf}
\newcommand{\be}{\begin{equation}}
\newcommand{\ee}{\end{equation}}
\newcommand{\ba}{\begin{eqnarray}}
\newcommand{\ea}{\end{eqnarray}}

\newcommand{\bi}{\begin{itemize}}
\newcommand{\ei}{\end{itemize}}
\newcommand{\bfi}{\begin{figure}
\epsfxsize=9cm \epsffile}
\newcommand{\efi}{\end{figure}}

\newcommand\aap{A\&A}                
\newcommand\aj{AJ}                   
\newcommand\apj{ApJ}                 
\newcommand\caa{Chinese Astron. Astrophys.} 
\newcommand\jcap{J.~Cosmology Astropart. Phys.} 
\newcommand\mnras{MNRAS}             
\newcommand\prd{Phys. Rev.~D}        

\title[Testing the anisotropy of cosmic acceleration]{Testing the anisotropy of cosmic acceleration from Pantheon supernovae sample}

\author[Sun \& Wang]{Z. Q. Sun$^{1}$, F. Y. Wang$^{1,2}$
\thanks{fayinwang@nju.edu.cn(FYW)}\\
$^{1}$School of Astronomy and Space Science, Nanjing University,
Nanjing
210093, China\\
$^{2}$Key Laboratory of Modern Astronomy and Astrophysics (Nanjing
University), Ministry of Education, Nanjing 210093, China }

\begin{document}
\maketitle
\begin{abstract}
In this paper, we study the anisotropy of cosmic acceleration the
using Pantheon sample, which includes 1048 spectroscopically
confirmed Type Ia supernovae (SNe Ia) covering the redshift range
$0.01 < z < 2.3$. In hemisphere comparison method, we find the
dipole direction is $(l = 37 \pm 40^{\circ}, b = 33 \pm 16^{\circ})$ with the maximum
anisotropy level of $\delta=0.136 {}^{+0.009}_{-0.005}$. From the dipole fitting method, we
find that the magnitude of anisotropy is $A = (3.7 {}^{+2.5}_{-3.7}) \times 10^{-4}$, and the direction of the dipole $(l = 329^{\circ}{}^{+ 101^{\circ}}_{-28^{\circ}}, b =
37^{\circ}{}^{+ 52^{\circ}}_{-21^{\circ}})$ in the galactic coordinate system. The result is weakly dependent
on redshift from the redshift tomography analysis.
The anisotropy is small and the isotropic cosmological model is an
excellent approximation.
\end{abstract}

\section{Introduction}
Observations of Type Ia supernovae (SNe Ia) have revealed that our Universe is
undergoing an era of accelerating expansion \citep{Perlmutter1999,Riess1998}. Observations from
cosmic microwave background (CMB; \citealt{Planck2016}), baryon acoustic oscillations
(BAO), clusters of galaxies, gamma-ray bursts (GRBs; \citealt{Wang2015}) and large-scale structure also support the accelerating
expansion. Therefore, the $\Lambda$-Cold Dark Matter ($\Lambda$CDM) model is widely accepted as
the standard model in modern cosmology. However, the $\Lambda$CDM
model also faces some challenges. For example, there exists a large
bulk-flow velocity at scales up to about 100h$^{-1}$ Mpc \citep{Lavaux2010}. The
``great cold spot" on CMB sky map \citep{Vielva2004},
alignment of polarization directions of quasars in large
scale \citep{Hutsemekers2005},
anisotropic distribution of gamma-ray bursts\citep{2016ChA&A..40...12G}
and spatial variation of the fine
structure constant \citep{King2012, Mariano2012} show that the Universe may be anisotropic.

SNe Ia have been widely used to study the possible anisotropy of
cosmic acceleration. \citet{Antoniou2010} firstly
used the hemisphere comparison method to search for a preferred axis
from the Union2 data set. They found that the expansion rate reaches its
maximum in a certain direction. \citet{Cai2011} found a similar
result in the $w$CDM model. \citet{Wang2014} firstly used SNe Ia
and GRBs to study the anisotropic expansion. They found the dipolar
anisotropy is more significant at low redshift from the redshift
tomography analysis. Meanwhile, several groups have applied the
hemisphere comparison method and dipole fitting method to search for
preferred direction of cosmic expansion \citep{Mariano2012, Cai2013, Zhao2013, Yang2014, 2015ApJ...810...47J, Jimenez2015, Chang2015, Lin2016a, Lin2016, Chang17}.
\citet{Sun2018} compared the results derived from
different SNe Ia samples and found that the anisotropic distribution of
SNe Ia coordinates can cause dipole directions and make dipole
magnitude larger. Although the magnitude of isotropy is small enough
such that the $\Lambda$CDM model is an excellent approximation, an
anisotropic cosmological model cannot be ruled out. Theoretically,
some anisotropic cosmological models have been proposed \citep{Campanelli2011, Mariano2012, Chang14, Wang2018}.

Recently, \citet{Scolnic2018} have compiled a larger SNe Ia sample
(called Pantheon) than JLA sample \citep{Jones2018}. Pantheon sample consists 1048 SNe
Ia covering the redshift range $0.01 < z < 2.3$. The coordinates of
SNe Ia in Pantheon sample distribute more uniformly than previous
samples, which is important to study the anisotropic expansion \citep{Sun2018}. In this paper, we investigate the anisotropy of
cosmic acceleration using the Pantheon sample. The hemisphere
comparison method and dipole fitting method are used.

The paper is organized as follows. In the next section, we introduce
the Pantheon sample and methods. The results are given in section \ref{sec:results}.
The summary is given in section \ref{sec:summary}.

\section{The Pantheon sample and methods}

\subsection{The Pantheon sample}
\begin{figure}
\includegraphics[width=1.0\columnwidth]{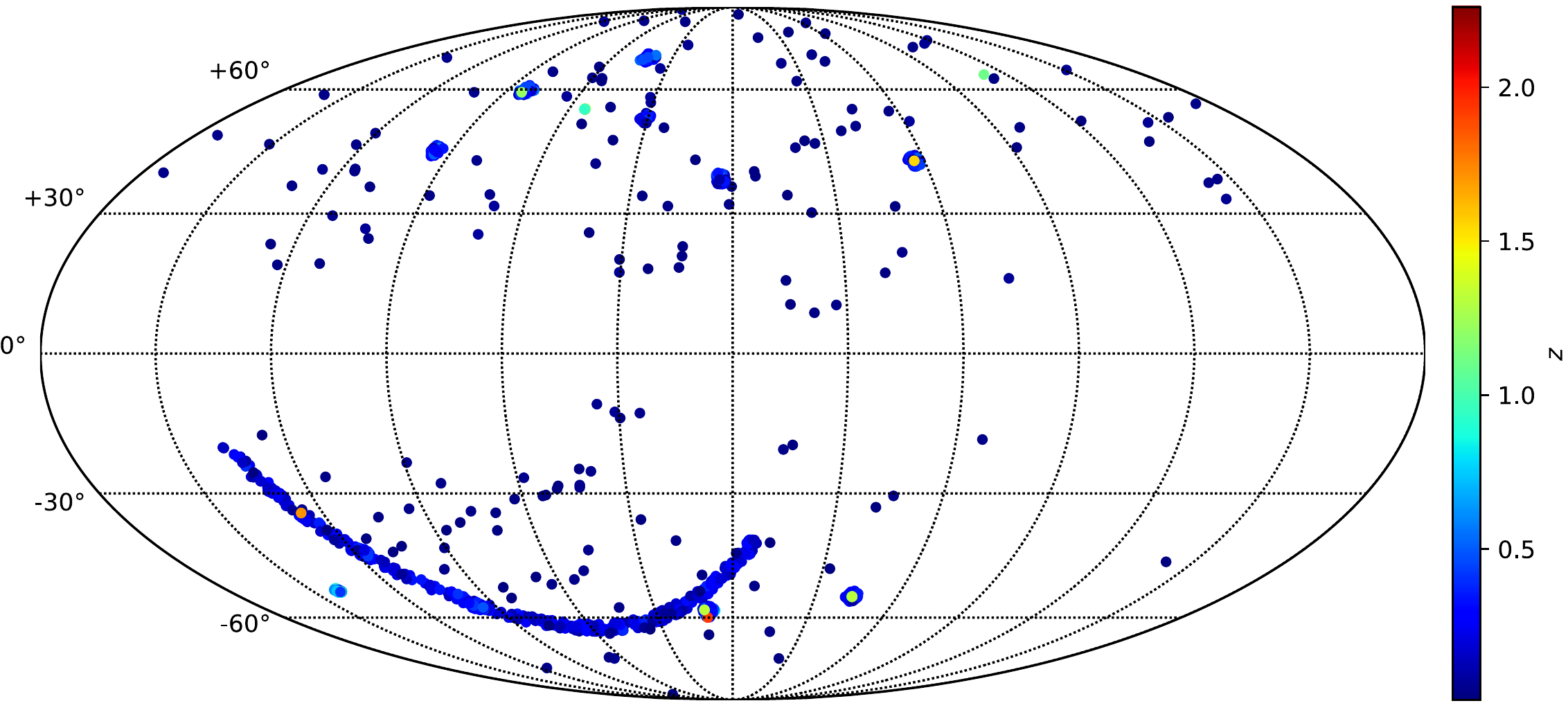}
\caption{The distribution of Pantheon sample in the galactic
coordinates. }\label{fig1}
\end{figure}

\citet{Scolnic2018} compiled the Pantheon sample consisting 1048
SNe Ia covering the redshift range $0.01 < z < 2.3$. This sample
contains 276 SNe Ia discovered by the Pan-STARRS1 Medium Deep
Survey, and SNe Ia from SDSS, SNLS, various low-$z$ and HST samples.
The distribution of SNe Ia in the galactic coordinates is shown in
Fig. \ref{fig1}. The luminosity distance is
\begin{equation}
D_\mathrm{L}(z)=(1+z)\int_{0}^{z}\frac{d{z}\rq{}}{E({z}\rq{})}.
\end{equation}
In the flat $\Lambda$CDM model, $E({z})$ is
\begin{equation}
E^2(z)=\Omega_\mathrm{m0}(1+z)^3+(1-\Omega_\mathrm{m0}),
\end{equation}
where $\Omega_\mathrm{m0}$ is the matter density.

The $\chi^{2}$ for SNe Ia is obtained by comparing theoretical
distance modulus
\begin{equation}
\mu_\mathrm{th}(z)=5\log_{10}\big(D_\mathrm{L}(z)\big)+\mu_{0},
\end{equation}
with
\begin{equation}
\mu_{0}=42.38-5\log_{10}h.
\end{equation}
The value of $\Omega_\mathrm{m0}$ is determined by minimizing the value of
$\chi^{2}$ with observed $\mu_\mathrm{obs}$
\begin{equation}
\chi_{\bf SN}^{2}(\Omega_\mathrm{m0}
,\mu_{0})=\sum_{i=1}^{1048}\frac{\Big(\mu_\mathrm{obs}(z_{i})-\mu_\mathrm{th}(\Omega_\mathrm{m0}
,\mu_{0},z_{i})\Big)^{2}}{\sigma_{\mu}^{2}(z_i)}.
\end{equation}
We can expand $\chi_{\bf SN}^{2}$ with respect to $\mu_{0}$
\citep{Nesseris2005}
\begin{equation}
\chi_{\bf
SN}^{2}=A-2\mu_{0}B+\mu_{0}^{2}C,\label{eq:expand}
\end{equation}
here
\begin{eqnarray*}
A & = & \sum_{i=1}^{1048}\frac{\big(\mu_\mathrm{obs}(z_{i})-\mu_\mathrm{th}(z_{i}, \mu_{0}=0)\big)^{2}}{\sigma_{\mu}^{2}(z_{i})},\\
B & = & \sum_{i=1}^{1048}\frac{\mu_\mathrm{obs}(z_{i})-\mu_\mathrm{th}(z_{i}, \mu_{0}=0)}{\sigma_{\mu}^{2}(z_{i})},\\
C & = &
\sum_{i=1}^{1048}\frac{1}{\sigma_{\mu}^{2}(z_{i})}.
\end{eqnarray*}
The value of Eq.~(\ref{eq:expand}) has a minimum for $\mu_0=B/C$ at
\begin{equation}
\widetilde{\chi}_{\bf SN}^{2}=\chi_{\bf SN, min}^{2}=A-B^{2}/C,
\end{equation}
which is not dependent on $\mu_{0}$.

\subsection{Hemisphere comparison method}

The hemisphere comparison method has been widely used in searching
for a preferred direction of cosmic expansion \citep{Antoniou2010, Yang2014}.
For estimating $\Omega_\mathrm{m0}$ in $\Lambda$CDM model, we can define \citep{Antoniou2010}
\begin{equation}
\label{eq1} \delta=\frac{\Delta
\Omega_\mathrm{m0}}{\bar\Omega_\mathrm{m0}}=\frac{\Omega_\mathrm{m0,u}-\Omega_\mathrm{m0,d}}{(\Omega_\mathrm{m0,u}+\Omega_\mathrm{m0,d})/2},
\end{equation}
where the subscripts $u$ and $d$ represent the best parameter
fitting value in the `up' and `down' hemispheres, respectively.
We search for axes that maximize $\delta$ by first evaluating $\delta$ on randomly
chosen axes, then using that axis with maximized $\delta$ as the initial value for the Nelder-Mead method.
the number of data points per hemisphere is approximate 500, we choose 500 axes in this
work. Since the hemisphere comparison approach is not pretty fine
and sensitive enough to particular types of anisotropy
\citep{Mariano2012}, it is just a rough estimation of the global
property.

\subsection{Dipole fitting method}
\citet{Mariano2012} firstly applied this method to
anisotropic study using SNe Ia. we define the
distance moduli with dipole $\boldsymbol{A}$ and monopole $B$ as
\begin{equation}
\tilde{\mu}_\mathrm{th} = \mu_\mathrm{th}(1 - \boldsymbol{A} \cdot
\hat{\boldsymbol{n}} + B),
\end{equation}
where $\tilde{\mu}_\mathrm{th}$ is the theoretical value of distance
modulus with dipolar direction dependence, and
$\hat{\boldsymbol{n}}$ is the unit vector pointing at the
corresponding SN Ia. $\hat{\boldsymbol{n}}$ can be expressed as
\begin{equation}
\hat{\boldsymbol{n}} = \cos (b) \cos (l) \hat{\boldsymbol{i}} + \cos
(b) \sin (l) \hat{\boldsymbol{j}} + \sin (b) \hat{\boldsymbol{k}}.
\end{equation}
Then the projection is
\begin{equation}
\boldsymbol{A} \cdot \hat{\boldsymbol{n}} = \cos (b) \cos (l)A_x +
\cos (b) \sin (l) A_y + \sin (b) A_z.
\end{equation}
The best-fitting dipole and monopole parameters can be derived by
minimizing $\chi^2_{SN}$.
The likelihood of the fitted parameters and the significance of
dipole magnitude are derived from Markov Chain Monte Carlo (MCMC)
sampling, similar to \citet{Sun2018}.

We construct three types of Monte Carlo samples, which are obtained with
distance moduli and coordinates replaced with different synthetic data while
keeping redshifts and other properties unchanged.
\begin{enumerate}
    \item For the first type of sample, $\mu_\mathrm{obs}$ is replaced with synthetic data subject to Gaussian distribution with $\mu_\mathrm{obs}$ as the mean value, and $\sigma$ as the standard deviation, where $\sigma$ is the error bar of $\mu_\mathrm{obs}$. Therefore, we assume the observational results of distance moduli are unbiased. The coordinates remain unchanged.
    \item For the second type of sample, $\mu_\mathrm{obs}$ is replaced with synthetic data generated by the same way as the first type, but with $\mu_\mathrm{obs}$ as the mean value, thus assuming underlying isotropy in the redshift-distance relation. The coordinates remain unchanged.
    \item In the third type of sample, $\mu_\mathrm{obs}$ is replaced with synthetic data generated by the same way as the second type, thus assuming underlying isotropy in the redshift-distance relation. In addition, coordinates are replaced with randomly generated coordinates uniformly distributed on the celestial sphere.
\end{enumerate}
For convenience, these three types of samples are called ``unbiased'', ``isotropic'' and
``random'' samples, respectively.
Likelihood distributions of the fitted parameters are approximated by the frequency
distribution of ``unbiased'' samples.
The significance of dipole magnitude is defined as the probability of
the best-fitting dipole magnitude being larger than that of an arbitrary data set in ``isotropic'' samples.

\section{Results}
\label{sec:results}
\begin{figure}
\includegraphics[width=\columnwidth]{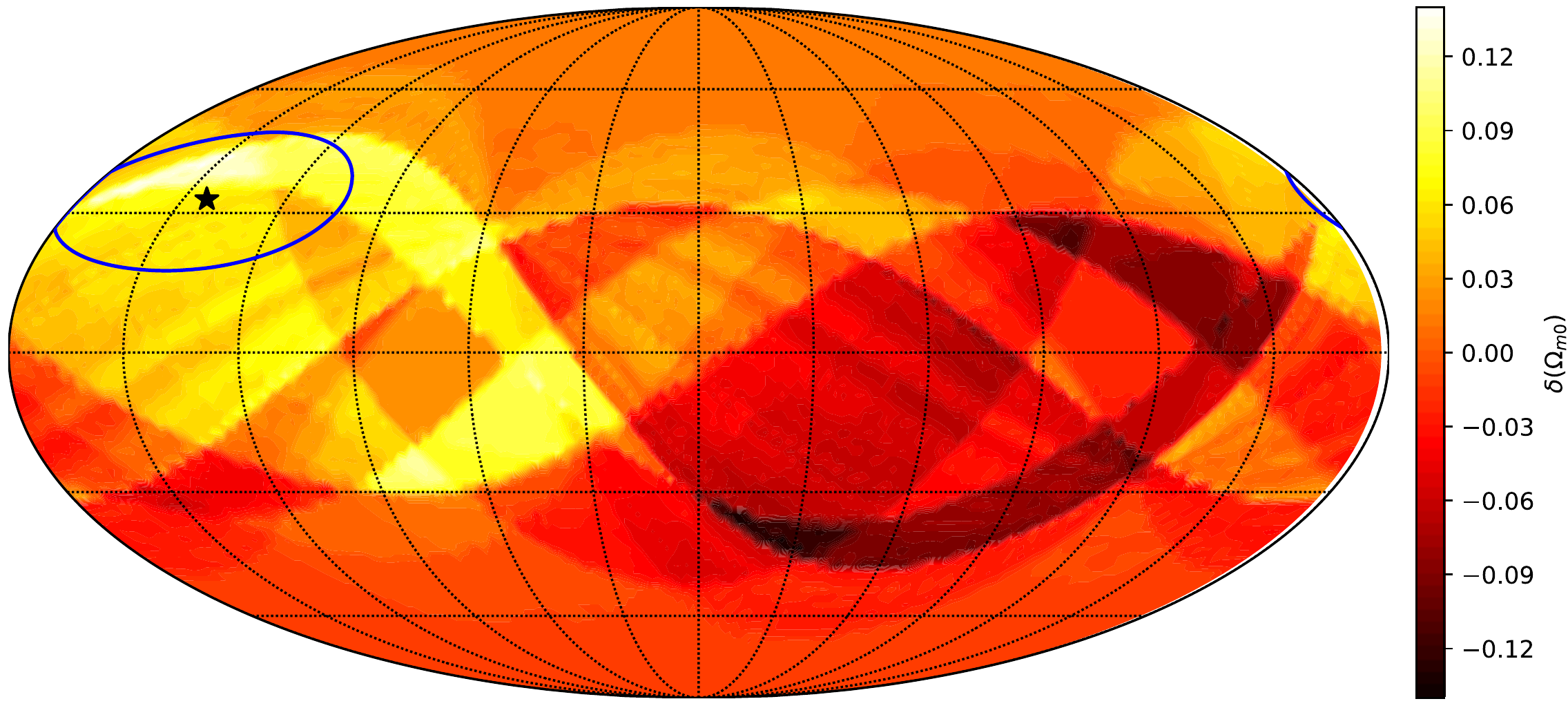}
\caption{The distribution of the hemispherical asymmetry $\delta =
\frac{\Omega_\mathrm{m0,u}-\Omega_\mathrm{m0,d}}{(\Omega_\mathrm{m0,u}+\Omega_\mathrm{m0,d})/2}$.
The star marks the direction of the largest $\delta$ , and the circle shows the error range.
}\label{fig:New_hemi}
\end{figure}

We apply the hemisphere comparison method using the Pantheon sample.
The direction of the largest $\delta$ is $(l = 37 \pm 40^{\circ}, b = 33 \pm 16^{\circ})$. The
maximum anisotropy level is $\delta_\mathrm{max}=0.136 {}^{+0.009}_{-0.005}$. Fig. \ref{fig:New_hemi} shows the
distribution of $\delta$. The star gives the direction of the
largest $\delta$, and the circle gives the error range.

\begin{figure}
\includegraphics[width=\columnwidth]{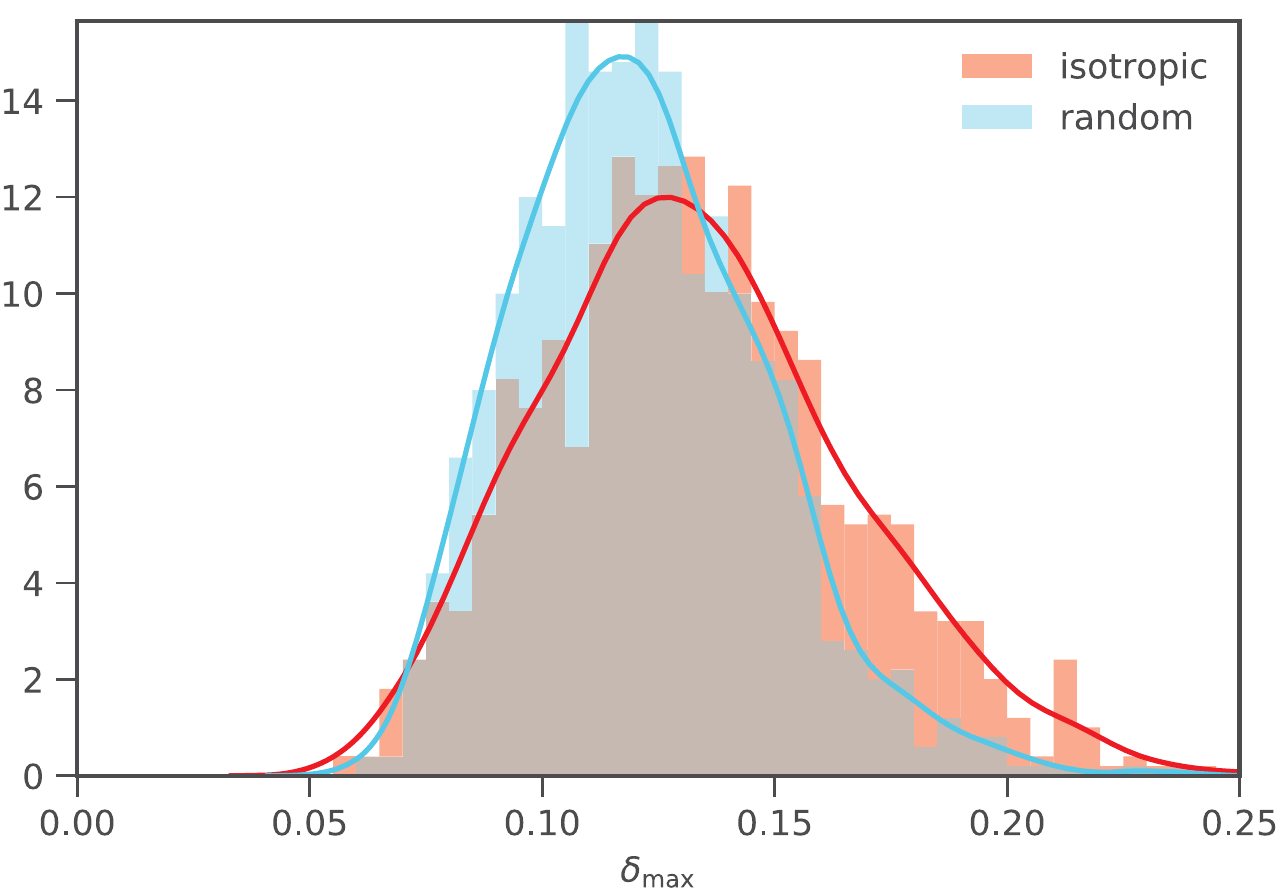}
\caption{
Histogram and kernel density plot of maximum anisotropy level $\delta_\mathrm{max}$ for ``isotropic'' and ``random'' samples.
}\label{fig:New_hemi_delta}
\end{figure}

We also apply the hemisphere comparison method for ``isotropic'' samples.
We get a maximum anisotropy level of $\delta_\mathrm{max}=0.133 {}^{+0.037}_{-0.036}$.
Using a similar definition as we introduced before, we can get the significance of
maximum anisotropy level $\delta_\mathrm{max}$ to be 83.6 per cent.
For ``random'' sample, we get a maximum anisotropy level of $\delta_\mathrm{max}=0.121 {}^{+0.028}_{-0.023}$.
The distribution of maximum anisotropy level $\delta_\mathrm{max}$ in ``isotropic'' and ``random''
samples are shown in Fig. \ref{fig:New_hemi_delta}.

Further, we explore the possible redshift dependence of the
anisotropy. We use a redshift tomography of the data and take the
same procedure as before for all the following redshift bins: 0-0.1,
0-0.2, 0-0.3, 0-0.4, 0-0.6, 0-2.3. The results of redshift
tomography analysis using hemisphere comparison method are shown in
Table \ref{table:hemisphere-tomography}. The redshift tomography
analysis shows that the preferred axes are all located in a
relatively small part of the North Galactic Hemisphere, which is consistent with the result of from the Constitution sample \citep{Sun2018}.

\begin{figure}
\includegraphics[width=\columnwidth]{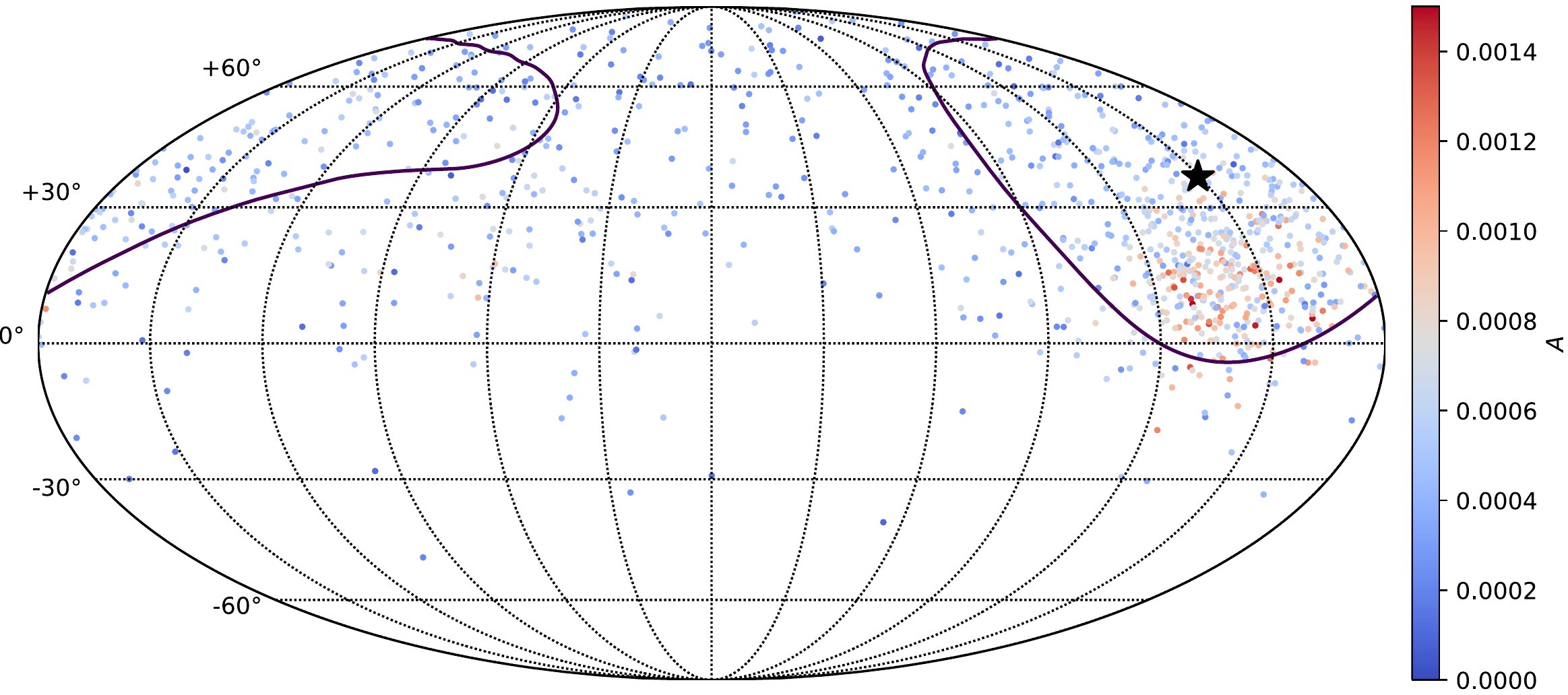}
\caption{Best-fitting dipole direction (star) and the 1$\sigma$ error
range of the Pantheon sample. Scatter points represent dipole
fitting results of ``unbiased'' samples.}
\label{fig:NewSamples}
\end{figure}

\begin{figure}
\includegraphics[width=\columnwidth]{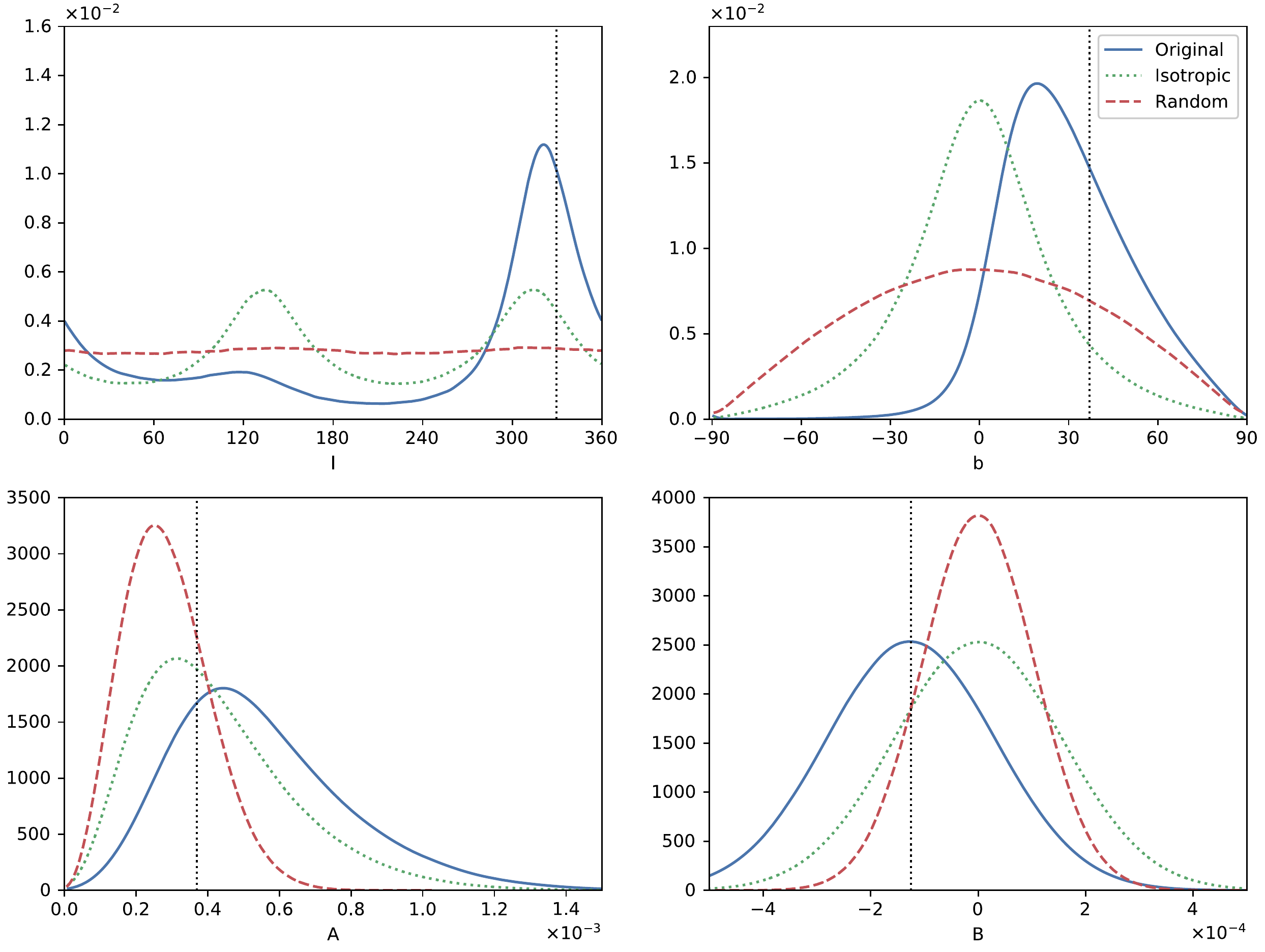}
\caption{The marginalized likelihoods for dipole $A$, monopole $B$ and $(l,b)$. Blue, green and orange lines show the results for the Pantheon sample, ``isotropic'' sample and ``random'' samples, respectively.
The black vertical lines represent best-fitting values.}
\label{fig:New_Hists}
\end{figure}

We also study the anisotropy of cosmic acceleration using the dipole
fitting method. We find the direction of the dipole is $(l =
329^{\circ}{}^{+ 101^{\circ}}_{-28^{\circ}}, b =
37^{\circ}{}^{+ 52^{\circ}}_{-21^{\circ}})$, which is shown as
the circle in Fig. \ref{fig:NewSamples}.
The scatter plot shows fitted dipole directions and magnitudes of
``unbiased'' samples.
The values of the dipole and monopole
magnitudes are $A = (3.7 {}^{+2.5}_{-3.7}) \times 10^{-4}$, and $B = (-1.3 \pm 1.6)
\times 10^{-4}$. The statistical significance of the dipole is
47.0 per cent, which is less than $1 \sigma$. Fig. \ref{fig:New_Hists} shows the best-fitting
results using the dipole fitting method.
For ``random'' samples, best-fitting dipole directions are evenly distributed
on the celestial sphere.
The dipole vectors $\boldsymbol{A}$ fit well with a spherically symmetric multivariate
Gaussian distribution centring at the origin. The mean value is $\mu = 0$ and standard deviation is
$\sigma = 1.8 \times 10^{-4}$ for each Cartesian components.
The monopole $B$ are also normally distributed, centring at $B = 0$. The standard
deviation is $\sigma = 1.0 \times 10^{-4}$.

\begin{figure}
\includegraphics[width=\columnwidth]{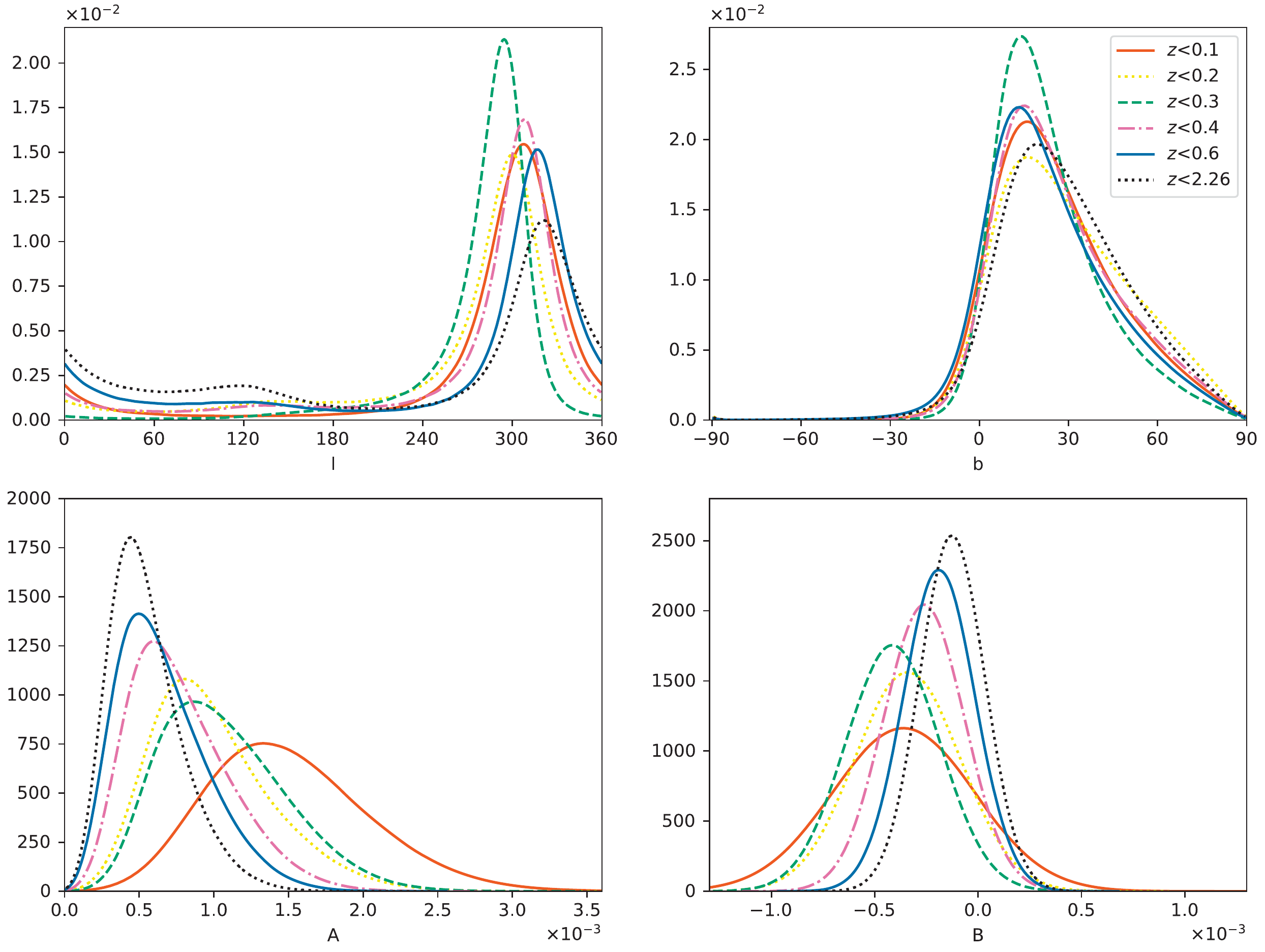}
\caption{Marginalized likelihoods of dipole $A$, monopole $B$ and
$(l,b)$ for different redshift ranges of the Pantheon sample.
}\label{fig:New_tomo_Hists}
\end{figure}

\begin{figure}
\includegraphics[width=\columnwidth]{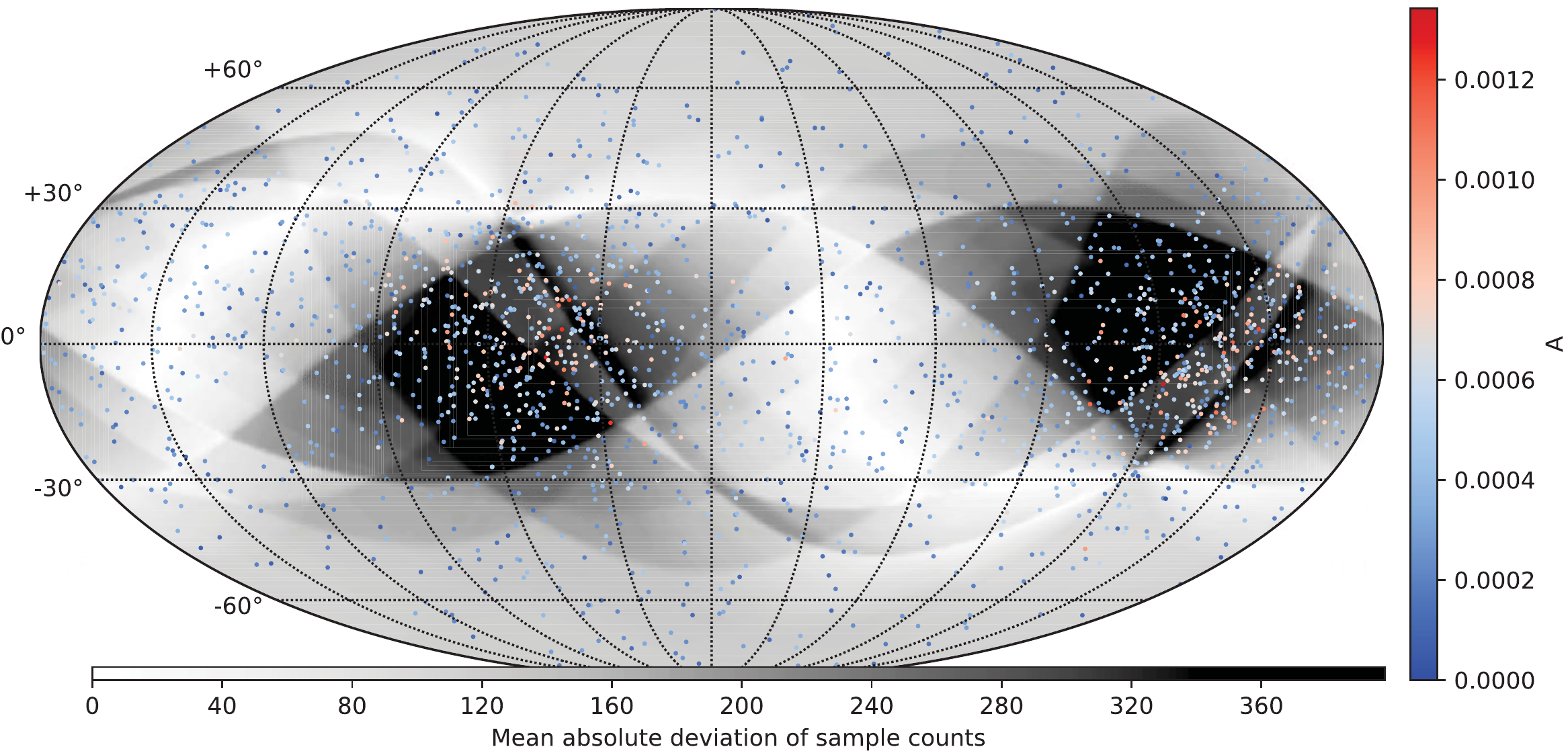}
\caption{Sample count and dipole distribution of the Pantheon samples.
The contour shows mean absolute deviations, which represent how far the sample density near a specific point differs from the average density.
}\label{fig:New_ISO_cnt}
\end{figure}

Similar to the hemisphere comparison method, we also perform the
redshift tomography analysis for this method. The redshift bins are
the same as above. The results of redshift tomography analysis using
dipole fitting method are shown in Table
\ref{table:dipole-tomography} and Fig. \ref{fig:New_tomo_Hists}. The redshift tomography
analysis shows that the preferred axes are all located in a
relatively small part of the Hemisphere, except for the first bin.
Meanwhile, the statistical significance of the dipole is similar.
To examine the anisotropy of the coordinate distribution of the Pantheon sample,
we calculate `sample counts' of data points in hemispheres centred at randomly given points.
In Fig. \ref{fig:New_ISO_cnt}, we plot `mean absolute deviations' of those sample counts,
along with dipole directions of ``isotropic'' samples.
We find that dipole directions ``isotropic'' samples tend to concentrate in places where sample counts deviate from the
average most.

\section{Summary}
\label{sec:summary}

From some observations, there seemingly exists some indications for
a cosmological preferred axis \citep{Perivolaropoulos2014}. If such a
cosmological preferred axis indeed exists, one has to consider an
anisotropic cosmological model as a realistic model, instead of the
standard cosmological model.

In this paper, we investigate the existence of anisotropy of the
Universe by employing the hemisphere comparison method and the
dipole fitting method using the Pantheon sample. For the hemisphere
method, we use the value of $\Omega_M$ to quantify the anisotropy
level. The dipole direction is $(l = 37 \pm 40^{\circ}, b = 33 \pm 16^{\circ})$.
The maximum anisotropy level is $\delta_\mathrm{max}=0.136 {}^{+0.009}_{-0.005}$.  For the dipole
fitting method, we find the direction of the dipole is $(l =
329^{\circ}{}^{+ 101^{\circ}}_{-28^{\circ}}, b =
37^{\circ}{}^{+ 52^{\circ}}_{-21^{\circ}})$. The values of the
dipole and monopole magnitudes are $A = (3.7 {}^{+2.5}_{-3.7}) \times 10^{-4}, B = (-1.3 \pm 1.6) \times 10^{-4}$. The statistical
significance of the dipole is 47.0 per cent, which is less than $1 \sigma$. Finally,
the redshift tomography analysis is used to explore the possible
redshift dependence of the anisotropy. From Tables 1 and 2, we find that the result is weakly dependent
on redshift.

\section*{Acknowledgements}
We thank the anonymous referee for constructive comments. We thank D. M. Scolnic for providing the Pantheon sample. This work
is supported by the National Basic Research Program of China (973
Program, grant No. 2014CB845800) and the National Natural Science
Foundation of China (grants 11422325 and 11373022), the Excellent Youth Foundation of Jiangsu Province
(BK20140016).


\newpage

\begin{table}
\centering
\caption{Redshift tomography results using hemisphere comparison
method.}
    \label{table:hemisphere-tomography}
    \begin{tabular}{ccrrr}
        \hline
        Redshift Range & Data Number &         $l$ &         $b$ & $\delta$ \\
        \hline
         $ z \leq$ 0.1 &         211 & 171$^\circ$ &  34$^\circ$ &     1.04 \\
         $ z \leq$ 0.2 &         411 & -47$^\circ$ &  37$^\circ$ &     0.54 \\
         $ z \leq$ 0.3 &         630 &  -9$^\circ$ &  50$^\circ$ &     0.38 \\
         $ z \leq$ 0.4 &         766 &  28$^\circ$ &  43$^\circ$ &     0.23 \\
         $ z \leq$ 0.6 &         886 &  16$^\circ$ &  36$^\circ$ &     0.18 \\
         $ z \leq$ 2.3 &        1048 &  37$^\circ$ &  33$^\circ$ &     0.14 \\
        \hline
    \end{tabular}
\end{table}

\begin{table}
\centering
\caption{Redshift tomography results using dipole fitting method.}
    \label{table:dipole-tomography}
    \begin{tabular}{ccrrrr}
        \hline
        Redshift Range  & Data Number &         $l$ &        $b$ &                  $A$ &                   $B$ \\
        \hline
          $ z \leq$ 0.1 &         211 & 307$^\circ$ & 25$^\circ$ & $1.3 \times 10^{-3}$ & $-3.6 \times 10^{-4} $ \\
          $ z \leq$ 0.2 &         411 & 296$^\circ$ & 30$^\circ$ & $7.7 \times 10^{-4}$ & $-3.4 \times 10^{-4} $ \\
          $ z \leq$ 0.3 &         630 & 289$^\circ$ & 21$^\circ$ & $9.5 \times 10^{-4}$ & $-4.1 \times 10^{-4} $ \\
          $ z \leq$ 0.4 &         766 & 306$^\circ$ & 26$^\circ$ & $6.3 \times 10^{-4}$ & $-2.6 \times 10^{-4} $ \\
          $ z \leq$ 0.6 &         886 & 320$^\circ$ & 24$^\circ$ & $5.1 \times 10^{-4}$ & $-1.9 \times 10^{-4} $ \\
          $ z \leq$ 2.3 &        1048 & 329$^\circ$ & 37$^\circ$ & $3.7 \times 10^{-4}$ & $-1.3 \times 10^{-4} $ \\
        \hline
    \end{tabular}
\end{table}

\end{document}